\documentstyle[11pt,newpasp,twoside,epsf]{article} 
\def\edcomment#1{\iffalse\marginpar{\raggedright\sl#1\/}\else\relax\fi} 
\reversemarginpar 

\begin{document} 
\title{Spectroscopy of Main Sequence Stars in Globular Clusters}

\author{Russell Cannon} 
\affil{Anglo-Australian Observatory, PO Box 296, Epping, NSW 2121,
Australia} 
\author{Gary Da Costa, John Norris, Laura Stanford}
\affil{Research School of Astronomy \& Astrophysics, Australian National
  University, Mount Stromlo Observatory, Cotter Road, Weston, ACT 2611, 
  Australia}
\author{Barry Croke}
\affil{Integrated Catchment Assessment \& Management Centre, and Centre for
  Resource \& Environmental Studies, Australian National University, Canberra
  ACT 0200, Australia}

\begin{abstract}
  Although globular clusters are generally chemically homogeneous, substantial
  abundance variations are sometimes seen even among unevolved main sequence
  stars, especially for the CNO group of elements.  Multi-object
  intermediate-dispersion spectroscopic systems are now being used to
  determine the patterns of abundance variations for large samples of stars,
  while high dispersion spectrographs on 8m-class telescopes are providing
  good spectra of individual faint stars.  In some circumstances the spectra
  of many similar stars can be combined to yield extremely high S/N spectra.
  The sample of clusters which has been studied remains small, but it seems
  that many of the more metal-rich clusters must have experienced some sort of
  self-enrichment, either in a proto-cluster stage, or through successive
  episodes of star formation or some other processes early in their lives.
  The metal-poor clusters provide equally clear evidence for internal mixing
  and dredge-up of processed material within evolving red giant stars.
\end{abstract}

\section{Introduction} 

This meeting is in honour of Ivan King, who has led the way in many aspects of
star cluster research.  Although he has not, so far as I am aware, done very
much spectroscopic work, we might be in better shape today if he had.  The
same principles which he and his colleagues have applied so effectively in the
careful analysis of HST astrometry and photometry apply equally in the field
of spectroscopy with ground-based telescopes.  We are now in the happy
position of being able to obtain good high dispersion spectra of tens of stars,
or intermediate dispersion spectra for hundreds of cluster stars, using
fibre-fed multi-object spectrographs on large telescopes.  As a result we are
no longer limited by the low signal-to-noise ratio (S/N) of individual spectra
or by small number statistics.  However, we are encountering new difficulties
and are often limited by systematic errors or calibration problems.  The
complexity of modern instruments also means that observers often have to treat
the control and data-analysis systems as `black boxes'.  The result is that
although we can obtain vast amounts of data rather quickly and easily, to
analyse them carefully is still as time-consuming as ever, and requires the
combined skills of teams of people with complementary expertise.  To finally
understand what we observe remains the hardest and slowest step.

The old assumption that all the stars in a given cluster have about the same
chemical composition and age remains a valid first approximation today, as
shown by the very well-defined sequences in the best colour-magnitude diagrams
(CMDs).  There is one glaring anomaly, $\omega$ Centauri, which certainly has
a wide spread in metallicity (by more than a factor of ten) and possibly a
wide range in star formation epoch; this cluster was the topic of an entire
meeting just one year ago (van Leeuwen, Hughes \& Piotto, 2002).  However, for
most well-observed globular clusters any spread in the abundance of iron or
other heavy elements is below the level of detectability.  Similarly, there is
no evidence for significant internal age spreads, although here the limits are
generally at the level of about 1-2~Gy or 10\% of the age of the clusters, so
there is scope for some initial spread in star formation.

The situation is however very different for some of the lighter
elements, especially C, N and O, and a few related species including
Na, Al and Mg.  Large variations in the strength of features due to CN
and CH were first noted more than 30 years ago (see e.g.\ the reviews
by Freeman \& Norris, 1981 and Kraft, 1994).  Initially these were
seen in bright red giants and were generally assumed to be due to the
dredge-up of processed material from the cores of the stars.  There is
good evidence that such processes do occur, especially in very
metal-poor clusters.  However, the CNO variations often persist down
on to the main sequence (m.s.) (e.g.\ Hesser, 1978), where neither the
requisite nuclear reactions nor the mixing mechanisms are supposed to
occur.  It seems that some stars contain substantial amounts of
material which have been processed through the CNO cycle in more
massive stars, and that there may have been an extended period of star
formation or some other processes of self-enrichment and accretion
during the early life of the clusters.

There are two main strands to recent work: observations of large samples of
stars using multi-object systems, and higher dispersion spectra for smaller
samples.  We concentrate more on the former, because others here (Gratton,
2002; several poster contributions) deal with the latest VLT and Keck results.
We have obtained spectra for a few hundred main sequence turn-off (MSTO) stars
in each of 47~Tuc, NGC~6752 and $\omega$~Cen, using the 2dF facility on the
AAT (Lewis et al, 2002).  Total exposure times were about 4 hours for each
sample of $\sim300$ stars, yielding spectra with S/N $\sim30$ per pixel and a
resolution of about 3\AA.  Two spectral regions were covered, blue spectra
extending from 3700\AA\ to 4800\AA\ and red spectra centred on the Na~D
lines at 5900\AA.  More details of the observations are given in Cannon et al
(2002).

It is convenient to group the clusters since those with similar metallicities
seem to show similar CNO abundance patterns.  However, it is not yet clear
whether this signifies a real metallicity dependence or merely reflects the
most easily observable effects, or indeed if it is simply an accident of the
small number of clusters studied so far.

\section{Relatively metal-rich clusters with [Fe/H]~$\sim$~--0.8}

\subsection{CN and CH in 47~Tuc}

The individual 47~Tuc blue spectra show obvious variations in the strengths of
the CH band near 4300\AA\ and the UV CN band near 3880\AA, although the
latter feature is harder to assess visually due to the converging Balmer
series of H lines and other features.  To provide a quantitative measure we
use simple band strength indices.  The result for one sample is shown in
Fig.~1.  There is a clear anti-correlation between the strengths of CN and CH
and the distribution appears to be bi-modal, at least in the more sensitive CN
index.  Both of these features are qualitatively the same as seen in the red
giants in 47~Tuc (Norris \& Freeman, 1979).  The bi-modality may be to some
extent an artefact of non-linearity in the formation of the molecular bands
(Langer, 1985) but synthetic spectra show that there is certainly a large
range in the abundances of C and N.

\begin{figure}
\plotfiddle{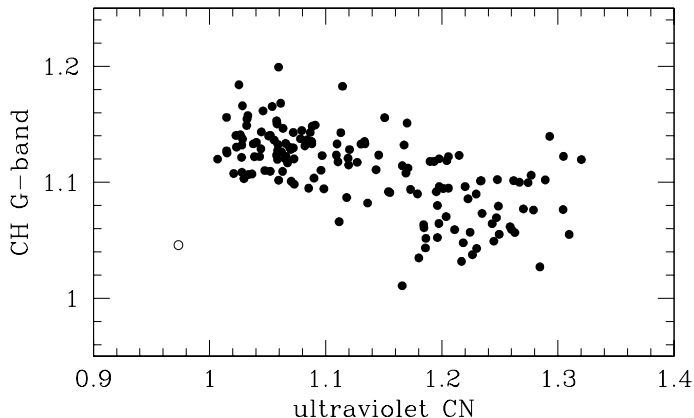}{6cm}{0}{50}{50}{-160}{-100}
\caption{The observed CH and CN indices for a sample of 140 stars
  around the m.s.\ turn-off in 47~Tuc.  CN and CH are anti-correlated, and the
  distribution in CN appears to be bi-modal.  The outlying star (open circle)
  is a radial velocity non-member.}
\end{figure}

We can split the sample in two according to CN strength and produce two very
high S/N spectra with about 200 hours of exposure each.  Figure~2 shows (a)
the mean spectra for some 50 CN-strong stars and 80 CN-weak stars, and (b) the
result of dividing one by the other.  The UV CN absorption shows up very
strongly in the quotient spectrum, while the anti-correlated CH G-band appears
in `emission', along with some broader features extending from 4200\AA\ to
4450\AA.  The CaII H \& K lines, which are the strongest features in the
original spectra, have cancelled almost perfectly, as have the Balmer series H
lines and other absorption lines due to Fe, CaI and other species.  This
accurate cancellation shows that the two samples of 47~Tuc stars must have
almost exactly the same mean effective temperature, surface gravity and heavy
element abundances.  It appears that the two samples of stars vary only in
their surface C and N abundances.

\begin{figure}
\plotfiddle{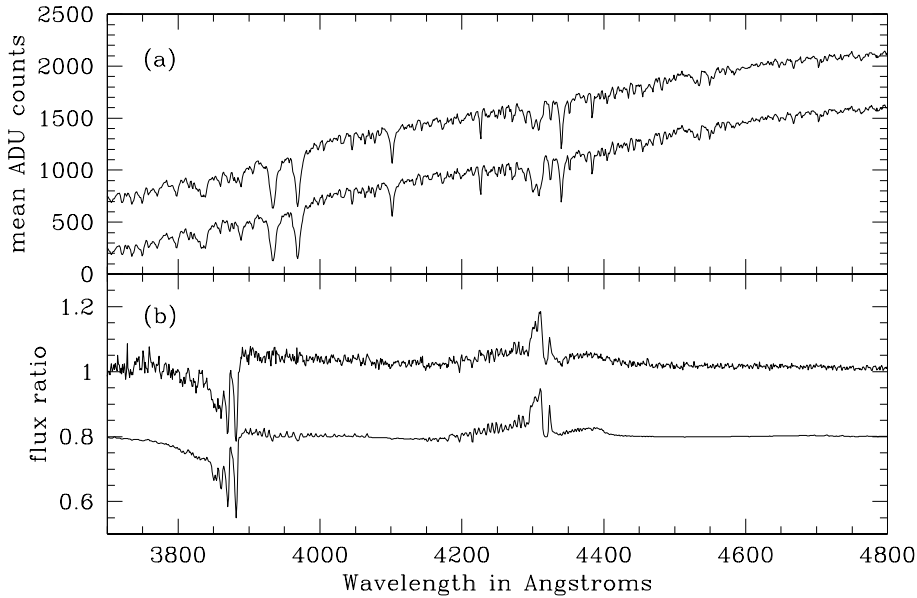}{8cm}{0}{70}{70}{-220}{-150}
\caption{(a) Comparison of the mean spectra for samples of 50 CN-strong
  stars near the 47~Tuc MSTO (upper trace, shifted by 500 counts) and 80
  CN-weak stars (lower); (b) the result of dividing the CN-strong spectrum by
  the CN-weak one (upper trace).  Only the features due to CN and CH remain
  when the spectra are divided.  The lower trace in (b) is a theoretical
  quotient spectrum (shifted down by 0.2 for clarity), obtained by dividing a
  matching pair of synthetic spectra.}
\end{figure}

In order to quantify the difference in C and N abundances we have made
synthetic spectra to match the 47~Tuc stars, using the {\it ssg} code of Bell
\& Gustafsson (1978) and MARCS model atmospheres (Gustafsson et al. 1975).  The
best fit is obtained with a pair of models having overall [Fe/H]~$=-0.7$, one
having [C/Fe] and [N/Fe] close to the solar values and the other with C
depleted by a factor of 2.3 and N enhanced by a factor of 10.  The synthetic
`quotient spectrum' is plotted in the lower panel of Fig.~2.  There is very
good agreement with the observed quotient spectrum, both in terms of the
overall strength of the molecular bands and their detailed structure.  The
numerical values of the C and N abundances are close to those derived
previously for much smaller (and lower S/N) samples of m.s.\ stars in 47~Tuc
by Briley et al. (1996) and Cannon et al. (1998), and to the abundances seen in
the evolved red giants.

Very recently Harbeck, Smith \& Grebel (2002) obtained VLT spectra for
a sample of 115 m.s.\ stars in 47~Tuc.  These extend more than 2 mag
fainter than the turn-off and show clearly the same CN dichotomy as
the 2dF data, with the strength of the molecular features increasing
markedly as the stars get fainter, presumably because the temperature
is decreasing.

\subsection{Sodium in 47~Tuc}

The individual red AAT 2dF spectra show few features apart from the
Na~D doublet at 5900\AA.  The result of combining the spectra of the
same 50 CN-strong stars as before is shown in Fig.~3.  Many fainter
absorption lines appear, most of them due to iron.  Some residual
night sky features are seen from strong [OI] emission lines at
5577\AA, 6300\AA\ and 6363\AA; these have not cancelled perfectly
because of their high intensity and small variations in the point
spread function of the images in the 2dF spectrographs.

\begin{figure}
\plotfiddle{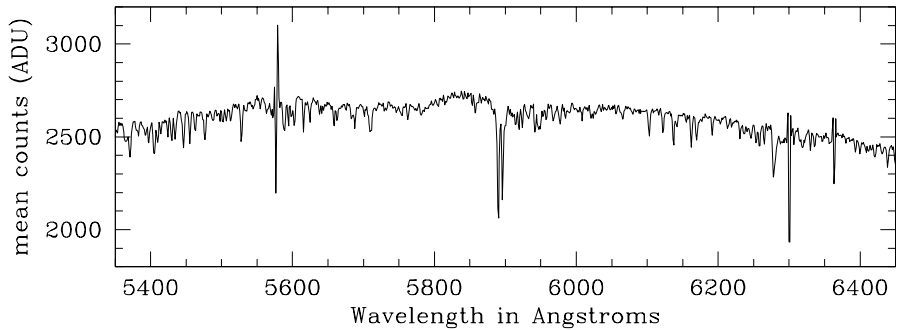}{5cm}{0}{70}{70}{-220}{-150}
\caption{The mean red 2dF spectrum for the sample of 50 CN-strong
stars at the MSTO in 47~Tuc.  Strong Na~D absorption is the dominant
feature but many weaker lines are present, mostly due to Fe.}
\end{figure}

When the mean red CN-strong spectrum is divided by the mean red
CN-weak spectrum, the only real features which remain are the Na~D
lines in absorption.  Evidently the Na absorption is correlated with
the N enhancement, as found by Cottrell \& Da Costa (1981) for red
giants, and by Briley et al. (1996) for a sample of six
m.s.\ stars in 47~Tuc.

\subsection{M71}

M71 has about the same metallicity as 47~Tuc but is much less massive
([Fe/H]~$=-0.8$; $M_{v}=-5.6$ cf $-9.4$ for 47~Tuc).  Briley \& Cohen (2001)
have obtained spectra for a sample of stars near the MSTO.  They find very
similar results to those reported here for 47~Tuc, i.e. a strongly bi-modal
distribution of CN band strengths, anti-correlated with CH.  This similarity
removes the previous worry that perhaps 47~Tuc is atypical, being one of the
most massive globular clusters and with a very concentrated core.  In another
recent result based on the same Keck spectra, Ramirez \& Cohen (2002) have
found that there are no measurable differences in the iron abundances of stars
across the CMD of M71.

\section{Intermediate-metallicity clusters, with [Fe/H]~$\sim$~--1.4}

\subsection{NGC~6752}

The red giants in NGC~6752, like those in 47~Tuc, show bi-modal CN
variations which are anti-correlated with CH (Norris et al, 1981).
However, the overall metallicity of NGC~6752 is about 4 times lower
than in 47~Tuc, so the molecular bands in MSTO stars are much weaker
because of both the lower abundance and the higher temperature; the
abundance effect is doubly severe in the case of CN.  Synthetic
spectra show that this is indeed the case: the bands have similar
structure to those in 47~Tuc, but the features are 10-20 times
weaker.

The same technique of combining spectra for many faint stars can be
used, but this time using the stronger CH G-band to divide the sample.
In NGC~6752 the quotient spectrum again shows UV CN absorption
anti-correlated with the CH feature, but the detection is rather
marginal and it is impossible to say whether the distribution is
bi-modal.

Independent evidence for main-sequence abundance variations in
NGC~6752 comes from recent high-dispersion VLT spectra for a small
sample of stars by Gratton et al. (2001).  They find that the oxygen
and sodium abundances are anti-correlated, while the Fe abundance
remains fixed at the same value across the CMD.

\subsection{M5}

Cohen, Briley \& Stetson (2002) have used LRIS on Keck to obtain
low-dispersion spectra for more than 40 stars near the base of the red giant
branch in M5, the largest uniformly observed sample of stars near the base of
the giant branch in any cluster.  They find strongly anti-correlated
variations in CN and CH, although no evidence for bi-modality.  By contrast,
Smith \& Norris (1983) found a strongly bi-modal CN distribution, again
anti-correlated with CH, for 29 brighter giants in M5.

\section{Metal-poor clusters, with [Fe/H]~$\sim$~--2.0}

The red giants in several low-metallicity clusters provide convincing evidence
of the dredge-up of internally processed material to the surface (e.g.\ Langer
et al, 1986 for M92; Briley et al, 1990, for NGC~6397).  It will probably be
impossible to detect the CN bands in turn-off stars in such metal-poor
clusters.  However, recent high resolution spectra taken with the ESO VLT for
m.s.\ stars in NGC~6397 ([Fe/H]~$\sim-2.1$) by Gratton et al. (2001) show that the O
and Na features remain constant in this cluster, as they do among the giant
stars, in contrast to the anti-correlated pattern seen in NGC 6752.  No
results for C or N have been reported as yet, but in another study of the same
set of 12 spectra, Bonifacio et al. (2002) find that all the stars have the same
Li abundance, at a level which is about equal to the primordial `Spite
plateau'.  On the other hand, Boesgaard et al. (1998) do find a significant
spread in Li abundance from Keck spectra for 7 m.s.\ stars in M92.

\section{The special case of $\omega$ Centauri}

While the majority of $\omega$~Cen stars are metal-poor with
[Fe/H]~$\sim-1.7$, it has red giant and m.s.\ members whose metallicity
extends to well above $-1.0$.  Recently, Lee et al. (1999) and Pancino et al.
(2000) discovered a very red, high metallicity component of the giant branch
in $\omega$~Cen.  If this comes from a substantially younger population of
stars (Hilker \& Richtler, 2000) then it should be easiest to detect the
different stellar populations in the vicinity of the subgiant branch, where
the isochrones are well separated.

Preliminary results from new AAT 2dF observations for a sample of 900 stars
redwards of the m.s., obtained in March 2002, show that faint red radial
velocity members of $\omega$~Cen do indeed exist, with $17.5 < V < 18.5$ and
$(B-V) > 0.8$, but the data analysis has not yet proceeded far enough to
determine the metallicities of these stars and hence to trace separate
isochrones.  It was already clear from earlier observations that there are
a number of very unusual stars lying to the red of the $\omega$~Cen main
sequence, including at least one with an extreme over-abundance of Sr by a
factor of about 100 (as reported by JEN at IAUXXIV, JD5, Manchester 2000).

\section{Interpretation and future work}

The last couple of years have seen a sudden expansion in our knowledge
of abundances in faint stars in globular clusters, coming both from
high resolution spectra with the new $8-10$m class telescopes and
from large samples of lower resolution spectra taken with multi-object
systems.  We still lack comprehensive homogeneous data for large
samples of stars in all parts of the CMD for any one cluster, but it
is now possible to make a few general statements.

\begin{itemize} 
  
\item N over-abundances correlate with C and O depletions and some Na
  enhancement.
  
\item The N abundance pattern often appears to be bi-modal, at least in the
  more metal-rich clusters, with comparable numbers of CN-strong and CN-weak
  stars.
  
\item The abundances of most heavy elements, such as Fe and Ca, show no
  detectable variations within each cluster (apart from $\omega$~Cen).
  
\item The C and N abundances needed to explain the m.s.\ results are
  approximately the same as those needed on the subgiant and giant branches.

\item Clusters with similar metallicities show similar abundance anomalies.

\end{itemize}

One other key datum is the evidence that field red giants in the Galaxy show
smaller abundance anomalies than similar stars in clusters (e.g.\ Langer et
al, 1992; Charbonnel \& Palacios here), which indicates that the CNO
variations are probably related to the cluster environment.

All these results suggest that the CNO and Na m.s.\ abundance variations
within clusters must be something to do with their formation and early
evolution.  They are apparently not due to internal nucleosynthesis and mixing
within the cluster stars themselves (although such effects also certainly
occur on the giant branch), nor can they be readily explained as a truly
primordial phenomenon (e.g.\ if clusters formed from the collisions of
separate gas clouds), since no variations are seen in Fe and Ca.

The similarity of the CNO and Na abundance patterns on the main
sequence, lower and upper giant branches indicates that the abundance
variations are inherent in the stars, involving at least the outer
30\% of their mass.  Any transient effects of a small amount of
surface pollution would be wiped out when the stars leave the m.s.\ and
develop fully convective envelopes at the base of the red giant
branch.

The simplest explanation is that there was some sort of self-enrichment
process within each cluster.  Whether this involved mechanisms of pollution,
accretion, mass exchange, binary star mergers or multiple generations of stars
is not yet clear.  Some or all of these processes may well have gone on during
the first few times $10^8$ years of the lives of the clusters, which being
$<5$\% of their current age would be very hard to detect in the CMDs today.

The observations imply that we are seeing the products of CNO-processed
material in m.s.\ stars in many clusters.  The most popular hypothesis for the
origin of this material is that it has come from intermediate-mass stars, with
masses in the range from say 2 to 5~ M$_{\sun}$, which do produce material
with the appropriate composition (e.g.\ Cottrell \& Da Costa, 1981; d'Antona,
Gratton \& Chieffi, 1983).  However, it is difficult to see how such material
could be so efficiently accreted, or why stars seem to accrete either a large
amount of material (more than 0.3~M$_{\sun}$) or almost none at all.  Thoul et
al. (2002; also poster at this conference) have revisited this problem and
find that sufficiently large amounts of gas can be accreted, at least in
massive concentrated clusters.  Alternatively, perhaps the answer lies in the
pre-supernovae stage for more massive stars, i.e. in stellar winds prior to
the Wolf-Rayet phase (R. Rood, private comm.), or even in supernovae
themselves, if the ejecta sometimes consist of only the outer layers where CNO
has been produced but not the heavier elements, as invoked to explain the
abundance patterns recently observed in some extremely metal-poor stars (e.g.\ 
Norris et al, 2002).  Or, it may be that binary stars, now being recognised
for their major effects in the dynamical evolution of clusters, may also be
the keys to understanding the abundance anomalies.  We need proper models of
the early lives of globular clusters, including the star formation process and
the full effects of gas flows and binary star interactions.

\end{document}